\documentclass[12pt,reqno]{amsart}
\usepackage{amsmath,amssymb,amsthm}

\newtheorem{lemma}{Lemma}
\newtheorem{proposition}{Proposition}

\newtheorem{corollary}{Corollary}

\theoremstyle{definition}
\newtheorem{remark}{Remark}
\newtheorem{example}{Example}
\newtheorem{definition}{Definition}

\newcommand{\Z}{\mathbb Z}

\newcommand{\N}{\mathbb N}
\newcommand{\C}{\mathbb C}
\newcommand{\R}{\mathbb R}
\newcommand{\K}{\mathcal K}
\newcommand{\A}{\mathcal A}
\newcommand{\B}{\mathcal B}
\newcommand{\torus}{\mathbb T}

\newcommand{\fii}{\varphi}
\newcommand{\hi}{\mathcal H}
\newcommand{\lh}{\mathcal{L(H)}}
\newcommand{\uh}{\mathcal{U(H)}}
\newcommand{\trh}{\mathcal{T(H)}}
\newcommand{\F}{\mathcal{F}}
\newcommand{\bor}[1]{\mathcal{B}({#1})}

\newcommand{\boto}{\mathcal B(\torus)}
\newcommand{\MT}{\textrm{M}(\torus)}
\newcommand{\supp}[1]{\textrm{supp }{#1}}

\newcommand{\M}{\mathcal{M}}

\newcommand{\ud}{\textrm{d}}
\newcommand{\ip}[2]{\langle {#1}|{#2}\rangle} 
\newcommand{\tr}{\mathrm{tr}}
\newcommand{\no}[1]{||{#1}||} 
\newcommand{\ketbra}[2]{|{#1}\rangle\langle {#2}|}

\newcommand{\dual}[2]{\langle {#1},{#2}\rangle}

\begin{document}

\title[Covariant fuzzy observables]{Covariant fuzzy observables and coarse-graining}

\author[Heinonen, Lahti, and Ylinen]{Teiko
  Heinonen$^{1,2,\dagger}$, Pekka
  Lahti$^{1,\ddagger}$, and Kari
  Ylinen$^{2,\S}$}

\address{$ ^1$Department of Physics, University of Turku, FIN-20014 Turku,
Finland}
\address{$ ^2$Department of Mathematics,
University of Turku, FIN-20014 Turku, Finland}
\address{$ ^\dagger$Electronic mail: temihe@utu.fi}
\address{$ ^\ddagger$Electronic mail: pekka.lahti@utu.fi}
\address{$ ^\S$Electronic mail: ylinen@utu.fi}
\maketitle

\begin{abstract}
A fuzzy observable is regarded as a smearing of a sharp
observable, and the structure of covariant fuzzy observables is
studied. It is shown that the covariant coarse-grainings of sharp
observables are exactly the covariant fuzzy observables. A necessary and
sufficient condition for a covariant fuzzy observable to be
informationally equivalent to the corresponding sharp observable
is given.
\end{abstract}

\section{Introduction}

The discovery of the proper mathematical formulation of the notion
of a physical quantity (observable) in quantum mechanics as a
normalized positive operator measure instead of the more
traditional spectral measure (normalized projection measure) has
provided, among many other advantages, a quantitative frame to
investigate various formulations of the idea of an imprecise
measurement of a physical quantity. In this paper we
study the structure of the so-called fuzzy observables, and, especially,
covariant fuzzy observables. 

The notion of a fuzzy observable will be formulated
as a smearing of a sharp observable (projection measure). The
definition of a fuzzy observable and some of its interpretations and motivations are given in Section \ref{fuzzy}. In Section
\ref{covariance} we recall some results concerning covariant
observables. In Section \ref{coarse} we show that another related
notion, the so-called coarse-graining, agrees with the notion of a fuzzy
observable in the covariant case. The structure of covariant fuzzy
observables is described. The norm-1-property of an observable (the
possibility that with a suitable preparation its measurement
outcome probabilities can be made arbitrarily close to one) as
well as the regularity (the Boolean structure of its range) are
operationally important properties. In Section \ref{properties} we
study consequences of these properties for fuzzy observables.
Since fuzzy observables are coarse-grainings of sharp observables,
the question arises under what conditions the state distinction
power of a fuzzy observable is the same as that of the
corresponding sharp observable. This will be answered in Section
\ref{equivalence}. In the final Section \ref{examples}
we shall illustrate some of the results of the paper in terms of
familiar examples.

\section{Fuzzy observables: definition and motivations}\label{fuzzy}

Let $\hi$ be a complex separable Hilbert space and $\lh$ the set
of bounded operators on $\hi$. Let $\Omega$ be a nonempty set and
$\F$ a $\sigma$-algebra of subsets of $\Omega$. We say that a set
function $E:\F\to\lh$ is an {\em operator measure}, if it is
$\sigma$-additive with respect to the strong (or \cite[p.
318]{DunSch1} equivalently, weak) operator topology. The operator
measure $E$ is {\em positive} if $E(X)\geq O$ for all $X\in\F$,
and {\em normalized} if $E(\Omega)=I$. If the normalized operator
measure $E$ is projection valued, that is, $E(X)^2=E(X)$ for all
$X\in\F$, it is a {\em spectral measure}. We call a positive
normalized operator measure $E:\F\to\lh$ an {\em observable} and a
spectral measure a {\em sharp observable}. For an observable
$E:\F\to\lh$ and a unit vector $\psi\in\hi$ we let $p_{\psi}^E$ denote
the probability measure on $\Omega$ defined by
$$
p_{\psi}^E(X)=\ip{\psi}{E(X)\psi},\ X\in\F.
$$
To distinguish sharp
observables from general observables, we reserve the letter $P$
for a spectral measure. If $\Omega$ is a topological space, we
denote by $\bor{\Omega}$ the Borel $\sigma$-algebra of $\Omega$.
For a selfadjoint operator $A$ in $\hi$ we let
$P^A:\bor{\R}\to\lh$ denote its spectral measure. In such a
situation we may also refer to $A$ as the sharp observable.

\begin{definition}\label{fuzzy:def}
Let $P:\F\to\lh$ be a sharp observable and consider a mapping
$\nu:\Omega\times\F\to [0,1]$, denoted also by
$(\omega,X)\mapsto\nu_{\omega}(X)$, with the following properties:
\begin{itemize}
\item[(i)] for every $\omega\in\Omega$, $\nu_{\omega}$ is a probability measure;
\item[(ii)] for every $X\in\F$, the mapping $\omega\mapsto \nu_{\omega}(X)$ is measurable.
\end{itemize}
The integral
\begin{equation}\label{fuzzy1}
E(X)=\int_{\Omega} \nu_{\omega}(X)\ \ud P(\omega),\quad X\in\F
\end{equation}
defines by the dominated convergence theorem a positive normalized operator measure. We call it a
\emph{fuzzy observable} with respect to $P$, and we say that the
mapping $\nu$ is a \emph{confidence measure}.
\end{definition}

There are at least two interpretations for the operator measure
$E$ defined in Equation (\ref{fuzzy1}), depending on whether one
wants to think of fuzziness as belonging to points or to events. We take
a brief look at these two viewpoints.

Since realistic measurements always have some unsharpness or
imprecision, one may think that the points of $\Omega$ are to be
replaced by probability distributions. We can write the sharp
observable $P$ in the form
\begin{equation}\label{spectral}
P(X)=\int_{\Omega} \delta_{\omega}(X)\ \ud P(\omega),
\end{equation}
where $\delta_{\omega}$ is the Dirac measure at a point
$\omega\in\Omega$. Comparing Equations (\ref{fuzzy1}) and
(\ref{spectral}), we notice that the fuzzy observable $E$ is
formed by replacing at every point $\omega\in\Omega$ the point
measure $\delta_{\omega}$ by the probability measure
$\nu_{\omega}$. Thus unsharpness related to a point $\omega$ is
quantified by a probability measure $\nu_{\omega}$. This is the
viewpoint e.g. in \cite{AliDoe}, \cite{AliPru} and \cite{Pru}.

Another interpretation arises from Zadeh's notion of a fuzzy set
and a fuzzy event, \cite{Zadeh1}, \cite{Zadeh2}. We recall that a
fuzzy set in $\Omega$ is a mapping $\widetilde{X}:\Omega\to[0,1]$,
and its values $\widetilde{X}(\omega)$ are called the membership
degrees of the points $\omega$. Ordinary sets are identified with
their characteristic functions. Let $\mu$ be a probability measure
on $\F$ . Following Zadeh, a fuzzy set is called a fuzzy event if
the corresponding function is measurable. The probability
$\mu(\tilde{X})$ of a fuzzy event $\widetilde{X}$ is defined as
the integral $\int_{\Omega}\widetilde{X}(\omega) \ud\mu(\omega)$.
Following this way of thought, it is natural to define
$P(\widetilde{X})$ as the operator
$$
P(\widetilde{X})=\int_{\Omega} \widetilde{X}(\omega)\ \ud
P(\omega).
$$
Let now $\nu$ be a confidence measure. For a fixed set $X\in\F$
the mapping $\omega\mapsto \nu_{\omega}(X)$ is a fuzzy event.
Thus $\nu$ can be seen as a function which maps measurable sets to
fuzzy events. If $\widetilde{X}$ is the fuzzy event $\omega\mapsto
\nu_{\omega}(X)$, Equation (\ref{fuzzy1}) can be written as
$$
E(X)=P(\widetilde{X}).
$$
We may thus conclude that in any vector state $\psi\in\hi,
\no{\psi}=1,$ the fuzzy observable $E$ gives for an event $X$ the
same probability as the sharp observable $P$ gives for a fuzzy
event $\widetilde{X}$.

To expand on the physical motivation of the present study, we
recall a typical situation of measurement theory, described by
the so-called standard measurement model \cite{BL1}, \cite{QTM}.
We denote by $\hi$ the Hilbert space of the system to be measured
and by $\mathcal{K}$ the Hilbert space of the measurement
apparatus. The aim is to measure a sharp observable $A$ of the
object system (a selfadjoint operator on $\hi$) using a
measurement interaction of the form
$$
U=e^{i\lambda A\otimes B},
$$
where $B$ is a sharp observable of the apparatus (a selfadjoint
operator on $\mathcal{K}$) and $\lambda\in\R$  a coupling
constant. Let $\phi\in\mathcal{K}$ be an initial vector state of
the apparatus and denote $\phi_{\lambda a}=e^{i\lambda aB}\phi$
for all $a\in\R$. Let $Z$ be the sharp pointer observable
(selfadjoint operator on $\mathcal K$) being applied. The probability
reproducibility condition
\begin{equation}
\ip{\fii}{E(X)\fii}= \ip{U(\fii\otimes\phi)}{ I\otimes
P^Z(X)U(\fii\otimes\phi)},
\end{equation}
required to hold for all vector states $\fii\in\hi$ and
$X\in\bor{\R}$, defines the actually measured observable
$E:\bor{\R}\to\lh$ of the object system, and a direct computation
shows that
\begin{equation}\label{standard}
E(X)=\int_{\R} \ip{\phi_{\lambda a}}{P^Z(X)\phi_{\lambda a}}\ \ud P^A(a).
\end{equation}
For a fixed $a$, the mapping $X\mapsto \ip{\phi_{\lambda
a}}{P^Z(X)\phi_{\lambda a}}$ is a probability measure and for a
fixed $X$, the mapping $a\mapsto \ip{\phi_{\lambda
a}}{P^Z(X)\phi_{\lambda a}}$ is continuous. Thus, the mapping
$(a,X)\mapsto \ip{\phi_{\lambda a}}{P^Z(X)\phi_{\lambda a}}$ is a
confidence measure. One concludes that the standard measurement
model $\langle\mathcal K, \phi,U,Z\rangle$ for a sharp observable
$A$ determines a fuzzy observable with respect to $A$.

\section{Covariant fuzzy observables}\label{covariance}

The behavior of an observable under appropriate transformations, like
space-time transformations, determines to a large extent the structure
of an observable. In such a relativistic approach, developed systematically by George W. Mackey,
one takes covariance as a defining property of an observable. In many
relevant cases (e.g. position, momentum, spin), although not in some
(e.g. phase, time), one has a unique covariant sharp observable, which
is also called the corresponding canonical observable. Our investigation in the rest
of this paper will concentrate on those covariant observables which
are fuzzy observables with respect to a canonical observable. In
this section we introduce further notation and recall some results,
which, if not otherwise stated, can be found e.g. in \cite{Fol1} and \cite{Fol2}.

Let $G$ be a locally compact topological group, which is Hausdorff
and satisfies the second axiom of countability. Then $G$ is also
metrizable and there exists a left invariant metric which induces
its topology (see e.g. \cite{Mon}). We fix a closed normal
(proper) subgroup $H$ of $G$, and from now on $\Omega$ denotes the quotient group $G/H$.
The groups $H$ and $\Omega$ are locally compact, Hausdorff and second
countable topological spaces in a natural way.
Moreover, the canonical projection
$$
\pi:G\to \Omega,\ g\mapsto gH
$$
is a continuous and open homomorphism.  We fix (arbitrary) left
invariant Haar measures of $G,H$ and $\Omega$ and denote them by
$\mu_G,\mu_H$, and $\mu_{\Omega}$. Let $d_G,d_H$, and $d_{\Omega}$
denote left invariant metrics. We denote by $(g,\omega)\mapsto
g\cdot\omega$ the action of $G$ on $\Omega$. If $g\in
G,\omega\in\Omega,X\subset \Omega$, we write $g\cdot X=\{g\cdot x
| x\in X\}$ and $\omega X=\{\omega x| x\in X\}$.

\begin{definition}
Let $U$ be a strongly continuous unitary representation of $G$ in
a Hilbert space $\hi$. We say that the observable
$E:\bor{\Omega}\to\lh$ is \emph{$G$-covariant} with respect to $U$
if
\begin{equation}
U(g)E(X)U(g)^*=E(g\cdot X)
\end{equation}
for all $g\in G$, $X\in\bor{\Omega}$. 
\end{definition}

In the sequel we will need the following result concerning the
structure of $G$-covariant observables. Let $E$ be a $G$-covariant observable. Then for any
$X\in\bor{\Omega}$,
\begin{equation}\label{nolla}
E(X)=O \textrm{ if and only if } \mu_{\Omega}(X)=0.
\end{equation}
For a proof see \cite[Prop. 8]{Norm1}.

We will assume that there exists a $G$-covariant sharp observable
$P:\bor{\Omega}\to\lh$ with respect to $U$. By the Imprimitivity
Theorem of Mackey \cite{Mackey} this is the case if and only if $U$ is
equivalent to a representation induced from some representation of
the subgroup $H$. Moreover, the pair $(U,P)$ is equivalent to the
canonical system of imprimitivity. This means that there is a separable
Hilbert space $\K$ such that we may (and always will) assume $\hi$ to be the space
$L^2(\Omega,\mu_{\Omega},\K)$ of $\K$-valued weakly measurable square-integrable
functions and $P(X)\psi=\chi_X\psi$ for all $X\in\bor{\Omega}$ and
$\psi\in\hi$ (for a convenient summary see e.g. \cite{AliAntGaz}). Throughout this article $(U,P)$ is a fixed canonical
system of imprimitivity, if not otherwise stated. If $E$ is a
$G$-covariant observable and fuzzy with respect to $P$, we call $E$ a
covariant fuzzy observable, for short. 

\begin{lemma}\label{confi}
Let $\rho:\bor{\Omega}\to [0,1]$ be a probability measure. The
mapping $\tilde{\rho}:\Omega\times\bor{\Omega}\to[0,1]$ defined by
$\tilde{\rho}_{\omega}(X)=\rho(\omega^{-1}X)$ is a confidence
measure.
\end{lemma}

\begin{proof}
The mapping $\tilde{\rho}$ clearly satisfies the condition (i) in
Definition \ref{fuzzy:def} so we concentrate on the condition
(ii). For any bounded Borel function $f:\Omega\to\C$, let us define the function $F_f:\Omega\to\C$ by 
$$
F_f(\omega)=\int f(\omega\eta)\ \ud\rho(\eta).
$$
In this notation $\rho(\omega^{-1}X)=F_{\chi_X}(\omega)$ so we have to show that the collection $\M=\{X\in\bor{\Omega}| F_{\chi_X} \textrm{ is
  measurable}\}$ coincides with
$\bor{\Omega}$. 

First, if $f$ belongs to the set $C_c(\Omega)$ of continuous
functions with compact support, then $f$ is left uniformly continuous and
$F_f$ is continuous. In particular, $F_f$ is measurable.

Let us then consider the case where $f=\chi_A$ for some open set
$A\subset\Omega$. As $\Omega$ is locally compact and second countable,
the set $A$ is $\sigma$-compact. Therefore, we may choose an
increasing sequence $(K_n)$ of compact sets such that $K_n\subset A$ and $\bigcup_n K_n=A$. Using
Urysohn's Lemma we find for each $n$ a function
$f_n\in C_c(\Omega)$ such that $0\leq f_n(\omega)\leq 1$ for all
$\omega\in\Omega$, $f_n(\omega)=1$ for all $\omega\in K_n$
and $f_n(\omega)=0$ for all $\omega\notin A$. The function $\chi_A$ is
the pointwise limit of the sequence $(f_n)$. By the dominated
convergence theorem 
\begin{eqnarray}\label{domi}
F_{\chi_A}(\omega) &=& \int \chi_A(\omega\eta)\ \ud \rho(\eta) = \int \lim_{n\to\infty} f_n(\omega\eta)\ \ud\rho(\eta)
\\
&=& \lim_{n\to\infty} \int f_n(\omega\eta)\ \ud\rho(\eta) =
\lim_{n\to\infty} F_{f_n}(\omega), \nonumber
\end{eqnarray}
and so $F_{\chi_A}$ is measurable.
Thus any open set belongs to $\M$. 

Denote by $\A$ the collection of those sets $X\subset\Omega$ for
which $\chi_X$ is of the form $\sum_{i=1}^n \epsilon_i \chi_{A_i}$ for some
open sets $A_i$ and numbers $\epsilon_i=\pm 1$. Clearly,
$\A$ is closed under complements and intersections and hence, it is an
algebra. As $F_{\sum_{i=1}^n \epsilon_i \chi_{A_i}}=\sum_{i=1}^n
\epsilon_i F_{\chi_{A_i}}$ and each $F_{\chi_{A_i}}$ is measurable, $\A\subset\M$.

An argument analogous to (\ref{domi}) shows that $\M$ is a monotone
  class. By the monotone class lemma \cite[p. 66]{Fol1} the monotone class $\M_0$ generated by $\A$ coincides with the $\sigma$-algebra $\B_0$
  generated by $\A$. Since $\B_0=\bor{\Omega}$ and
  $\M_0\subset\M\subset\bor{\Omega}$ we have $\M=\bor{\Omega}$.  
\end{proof}

\begin{proposition}\label{covar:prop}
Let $\rho:\bor{\Omega}\to [0,1]$ be a probability measure. Then the
observable $E_{\rho}$ defined by
\begin{equation}\label{fuzzy2}
E_{\rho}(X)=\int_{\Omega}\rho(\omega^{-1}X)\ \ud P(\omega),\quad
X\in\bor{\Omega},
\end{equation}
is a covariant fuzzy observable. Moreover, for any unit vector $\psi\in\hi$ the
probability measure $p^{E_{\rho}}_{\psi}$ takes the form
\begin{equation}\label{conv1}
p_{\psi}^{E_\rho}=p_{\psi}^P\ast\rho,
\end{equation}
where $p_{\psi}^P\ast\rho$ is the convolution of the measures
$p_{\psi}^P$ and $\rho$.
\end{proposition}

\begin{proof}
It follows directly from Lemma \ref{confi} that $E_{\rho}$ is a fuzzy
observable with respect to $P$. Let $g\in G$, $X\in\bor{\Omega}$, and
$\psi\in\hi$. Since $P$ is $G$-covariant, we have
\begin{eqnarray*}
\ip{\psi}{U(g)E_{\rho}(X)U(g)^*\psi} &=&
\int_{\Omega}\rho(\omega^{-1}X)\ \ud p^P_{U^*(g)\psi}(\omega) \\
&=& \int_{\Omega}\rho((g^{-1}\cdot\omega)^{-1}X)\ \ud p^P_{\psi}(\omega)
\\
&=& \int_{\Omega}\rho(\omega^{-1}(g\cdot X))\ \ud p^P_{\psi}(\omega)\\
&=& \ip{\psi}{E_{\rho}(g\cdot X)\psi}.
\end{eqnarray*}
Thus $E$ is $G$-covariant.

The last claim follows from the fact that for any
$X\in\bor{\Omega}$,
\begin{equation*}
p^{E_\rho}_{\psi}(X) = \int_{\Omega} \rho (\omega^{-1}X)\ \ud
p_{\psi}^P(\omega) = \int_{\Omega} \int_{\Omega} \chi_X(\omega\eta)\
\ud\rho (\eta) \ud p_{\psi}^P(\omega).
\end{equation*}
\end{proof}

The Dirac measure $\delta_{\omega}$ at a point
$\omega\in\Omega$ yields a fuzzy observable $E_{\delta_{\omega}}$,
which is the translated spectral measure
$$
X\mapsto P(X\omega^{-1}),
$$
also denoted by $P_{\omega}$. This differ from the canonical
spectral measure $P$ only by 'the choice of the origin'. Equation (\ref{fuzzy2}) can be written in the form
\begin{equation}\label{fuzzy3}
E_{\rho}(X)=\int_{\Omega} P(X\omega^{-1})\ \ud
\rho(\omega)=\int_{\Omega} P_{\omega}(X)\ \ud \rho(\omega) .
\end{equation}
Thus the operators $E_{\rho}(X)$ are weighted means of the
translated projections $P_{\omega}(X),\omega\in\Omega$.

\begin{example}
Let $\rho$ be a probability measure $\rho$ with a finite support
$\supp(\rho)=\{\omega_1,\omega_2,\ldots,\omega_n\}$ and
$\rho(\{\omega_i\})=\lambda_i$, $i=1,\ldots,n$. Equation
(\ref{fuzzy3}) becomes
\begin{equation}\label{fuzzy4}
E_{\rho}(X)=\sum_{i=1}^n \lambda_i P_{\omega_i}(X).
\end{equation}
As $\sum_{i=1}^n \lambda_i=1$, we see that the fuzzy observable
$E_{\rho}$ is a convex combination of the sharp observables
$P_{\omega_1},\ldots,P_{\omega_n}$. Therefore, we may think of
$E_{\rho}$ as a mixture of translated sharp observables.
\end{example}

\begin{example}\label{covar:ex}
Let $\rho$ be a probability measure and $\tilde{\rho}$ the corresponding
confidence measure (defined in Lemma \ref{confi}). Assume that $\rho$ is absolutely continuous with respect
to the Haar measure $\mu_{\Omega}$ and let $f$ be the corresponding
Radon-Nikod\' ym
derivative. Define the function $\check{f}$ by
$\check{f}(\omega)=f(\omega^{-1})$. For any set $X\in\bor{\Omega}$ the
confidence measure $\tilde{\rho}$ defines a fuzzy event
$\widetilde{X}$, as in Section \ref{fuzzy}. The fuzzy event
$\widetilde{X}$ can be written simply as a convolution of
the characteristic function $\chi_X$ and the function $\check{f}$, since 
\begin{eqnarray*}
\widetilde{X}(\omega) &=& \tilde{\rho}_{\omega}(X)=\rho(\omega^{-1}X)
= \int \chi_{\omega^{-1}X}(\eta)f(\eta)\ \ud\mu_{\Omega}(\eta)\\
&=& \left( \chi_X \ast \check{f} \right)(\omega).  
\end{eqnarray*} 
\end{example}

\section{Coarse-graining}\label{coarse}

Let $\trh$ be the set of all trace-class operators on $\hi$ and
$M(\Omega)$ the set of all complex Radon measures on $\Omega$. An
observable $E:\bor{\Omega}\to\lh$ defines a linear mapping
\begin{equation}
V_E:\trh\to M(\Omega),\ V_E(T)(X)=\tr [TE(X)].
\end{equation}
In particular, if $T$ is a state, i.e. a positive trace one
operator, then $V_E(T)$ is a probability measure. Moreover,
if $T$ is a vector state $T=\ketbra{\psi}{\psi}$, then
$V_E(T)=p_{\psi}^E$. The mapping $V_E$ characterizes the observable
$E$ completely in the sense that $V_E=V_{E'}$ implies $E=E'$.

\begin{definition}
Let $E_1$ and $E_2$ be observables from $\bor{\Omega}$ to $\lh$ and $V_{E_1},V_{E_2}$ the corresponding mappings. We say that the observable $E_2$ is
\emph{coarser} than $E_1$, if there exists a
linear map $W:M(\Omega)\to M(\Omega)$, such that
\begin{itemize}
\item[(i)] for any probability measure $m\in M(\Omega)$, $W(m)$ is
a probability measure;
\item[(ii)] $V_{E_2}$ is the composite mapping of $V_{E_1}$ and $W$, that is, $V_{E_2}=W\circ V_{E_1}$.
\end{itemize}
If $E_2$ is coarser than $E_1$ we also say that $E_2$ is a
\emph{coarse-graining} of $E_1$.
\end{definition}

Using the Jordan decomposition of a measure one infers from (i)
that $W$ is a bounded linear map.

If $E_2$ is a coarse-graining of $E_1$, for any two
states $S,T\in\trh_1^+$ the following implication holds:
\begin{equation}
V_{E_1}(S)=V_{E_1}(T)\ \Rightarrow\
V_{E_2}(S)=V_{E_2}(T).
\end{equation}
This means that the state distinction power of $E_2$ does not exceed
that of $E_1$.

\begin{example}\label{coarse:ex}
Fuzzy observables are coarser than the corresponding sharp
observable. Indeed, any confidence measure
$\nu:\Omega\times\bor{\Omega}\to [0,1]$ induces a linear mapping
$$
W_{\nu}:M(\Omega)\to M(\Omega),\
W_{\nu}(m)(X)=\int_{\Omega}\nu_{\omega}(X)\ \ud m(\omega).
$$
The observable $E$ corresponding to the map $W_{\nu}\circ V_P$ is
the fuzzy observable given by equation (\ref{fuzzy1}).
\end{example}

If $E$ is a $G$-covariant observable, equation (\ref{nolla})
implies that for any $T\in\trh$, the measure $V_E(T)$ is
absolutely continuous with respect to $\mu_{\Omega}$. Thus, we may
identify $V_E(T)$ with its Radon-Nikod\' ym derivative with respect to
$\mu_{\Omega}$ and hence the mapping $V_E$ with the corresponding mapping from $\trh$ to
$L^1(\Omega)$. For our subsequent needs we state and give a proof of the following known result. 

\begin{lemma}
The mapping $V_P:\trh\to L^1(\Omega)$ corresponding to the canonical
observable $P$ is surjective.
\end{lemma}

\begin{proof}
Let $f$ be a function in
$L^1(\Omega)$ and write it as $f=f_1-f_2+i(f_3-f_4)$, where $f_n,
n=1,2,3,4,$ are nonnegative functions and belong to $L^1(\Omega)$.
Fix a unit vector $\varphi\in\K$ and for each $n=1,2,3,4$ denote
$\psi_n(\omega)=\sqrt{f_n(\omega)}\varphi$. Then $\psi_n$ belongs
to $L^2(\Omega,\mu_{\Omega},\K)$ and
$V_P(\ketbra{\psi_n}{\psi_n})=f_n$. The operator
$$
T=\ketbra{\psi_1}{\psi_1}-\ketbra{\psi_2}{\psi_2}+i\ketbra{\psi_3}{\psi_3}-i\ketbra{\psi_4}{\psi_4}
$$
is a trace-class operator and $V_P(T)=f$.
\end{proof}

\begin{proposition}\label{coarse1}
Let $E$ be a $G$-covariant observable. If $E$ is coarser than $P$, then $E=E_{\rho}$ for some probability measure $\rho:\bor{\Omega}\to[0,1]$.
\end{proposition}

\begin{proof}
Assume that $E$ is coarser than $P$. By definition there
exists a linear map $W:L^1(\Omega)\to L^1(\Omega)$ such that
\begin{equation}\label{W}
V_E=W\circ V_P.
\end{equation}
For any $\xi\in \Omega$ and $f\in L^1(\Omega)$, we define
$(\Lambda_{\xi}f)(\omega)=f(\xi\omega)$. Then
$\Lambda_{\xi}:L^1(\Omega)\to L^1(\Omega)$ is a linear mapping. We
will show that for all $\xi\in\Omega$,
\begin{equation}\label{Lambda1}
\Lambda_{\xi}\circ W=W\circ \Lambda_{\xi}.
\end{equation}

Together with a result of Wendel \cite[Theorem 1]{Wendel} as
formulated in \cite[p. 376]{HR} this then implies that $W$ is of
the form $W(f)=f\ast\rho$ for some probability measure $\rho$. By
equation (\ref{conv1}) this means that $E=E_{\rho}$.

To prove (\ref{Lambda1}), fix $\xi\in\Omega$ and choose $g\in G$
such that $\pi(g)=\xi$. Since $E$ is covariant, we get
\begin{equation}\label{Lambda2}
\Lambda_{\xi}V_E(T)=V_E(U(g)^*TU(g))
\end{equation}
for all $T\in\trh$. The same is true for $V_P$, that is,
\begin{equation}\label{Lambda3}
\Lambda_{\xi}V_P(T)=V_P(U(g)^*TU(g)).
\end{equation}
Comparing equations (\ref{W}), (\ref{Lambda2}) and
(\ref{Lambda3}), we get
\begin{equation}\label{Lambda4}
\Lambda_{\xi}\circ W\circ V_P=W\circ \Lambda_{\xi}\circ V_P.
\end{equation}
As the mapping $V_P$ is surjective, equation (\ref{Lambda1})
follows from (\ref{Lambda4}).
\end{proof}

\begin{corollary}\label{cor}
Let $E$ be a $G$-covariant observable. The following are equivalent:
\begin{itemize}
\item[(i)] $E$ is a fuzzy observable with respect to $P$;
\item[(ii)] $E$ is coarser than $P$;
\item[(iii)] There is a probability measure $\rho$ such that
  $E=E_{\rho}$.
\end{itemize}
\end{corollary}

\begin{proof}
In Example \ref{coarse:ex} it was shown that (i) implies (ii). By
Proposition \ref{coarse1} (ii) implies (iii). That (iii) implies (i)
was shown in Proposition \ref{covar:prop}.
\end{proof}

\section{Properties of covariant fuzzy observables}\label{properties}

In this section we study the norm-1-property as well as the
regularity of a covariant fuzzy observable and we find a necessary
and sufficient condition for a fuzzy observable to be
informationally equivalent to its sharp counterpart.

\subsection{The norm-1-property}

Sharp observables $P$, i.e. spectral measures,  have the property that
the norm of any $P(X)\ne O$ is one. Spectral measures are not the
only observables having this property, called the norm-1-property
in \cite{Norm1}. Some examples are given in \cite{Norm1}. If an
observable $E$ has the norm-1-property, then for each $E(X)\ne O$
there is a sequence of vector states $(\fii_n)$ such that
$\lim_{n\to\infty}p^E_{\fii_n}(X)=1$. The following proposition
shows that among the covariant fuzzy observables only the
translated spectral measures have the norm-1-property.

\begin{proposition}\label{norm}
Let $E$ be a $G$-covariant observable which is fuzzy with respect
to $P$. Then $E$ has the norm-1-property if and only if
$E=P_{\omega}$ for some $\omega\in \Omega$.
\end{proposition}

\begin{proof}
If $E$ is a spectral measure, then it has the norm-1-property.
Assume now that $\rho$ is a probability measure on $\Omega$ and
$E$ is defined by equation (\ref{fuzzy2}). Then $E=P_{\omega}$ if
and only if $\rho=\delta_{\omega}$. Therefore it is enough to show
that if $E$ has the norm-1-property, then $\supp(\rho)$ is a one
point set.

For any $X\in\bor{\Omega}$,
$$
\no{E(X)}=\no{\int_{\Omega}\rho(\omega^{-1}X)\ud P(\omega)}\leq
\no{P(X)}\cdot \sup_{\omega\in\Omega}\rho(\omega^{-1}X) \leq
\sup_{\omega\in\Omega}\rho(\omega X).
$$
Thus $E$ cannot have the norm-1-property if
$\sup_{\omega\in\Omega} \rho(\omega X)<1$ for some
$X\in\bor{\Omega},\ \mu_{\Omega}(X)>0$. (By equation
(\ref{nolla}), $E(X)\neq O$ if and only if $\mu_{\Omega}(X)\neq
0$.)

Assume now that there are at least two points $x$ and $y$ in
$\supp(\rho)$ and denote $r=d_{\Omega}(x,y)$. Then the sets
$B(x;\frac{r}{5})$ and $B(y;\frac{r}{5})$ are disjoint and have
positive $\rho$-measure. Denote $m=\min
\{\rho(B(x;\frac{r}{5})),\rho(B(y;\frac{r}{5}))\}$. Because the
metric $d_{\Omega}$ is left invariant, we have $\omega
B(x;\frac{r}{5})=B(\omega x;\frac{r}{5})$ for any
$\omega\in\Omega$. Moreover, for any $\omega\in\Omega$, we have
$B(\omega x;\frac{r}{5})\cap B(x;\frac{r}{5})=\emptyset$ or
$B(\omega x;\frac{r}{5})\cap B(y;\frac{r}{5})=\emptyset$. This
implies that $\rho(\omega B(x;\frac{r}{5}))\leq 1-m$ for all
$\omega\in\Omega$ so that $\sup_{\omega\in\Omega}\rho(\omega
B(x;\frac{r}{5}))<1$.
\end{proof}

\subsection{Regularity}

An observable $E$ is called {\em regular}, if for each $O\ne
E(X)\ne I$, the spectrum of $E(X)$ extends both below and above
$\frac 12$. For such an observable, for each (nontrivial) $E(X)$
there are vector states $\fii$ such that $p^E_\fii(X)>1/2$ and
$p^E_\fii(X')<1/2$, a property which resembles the fact that for
each (nontrivial) projection $P(X)$ there is a $\fii$ such that
$p^P_\fii(X)=1$ and $p^P_\fii(X')=0$. If an observable $E$ is
regular, then its range $\{E(X)|X\in\F\}$ is a Boolean system with
respect to the natural order and complement of the set of effects
restricted to the range \cite{DvurPul}.

Clearly, an observable having the norm-1-property is also regular.
Next we give an example of covariant fuzzy observable which is
regular and which is not a spectral measure.

\begin{example}
Let $\omega_1,\omega_2\in\Omega$ be distinct points. Let $E$ be a proper
convex combination of the translated spectral measures $P_{\omega_1}$
and $P_{\omega_2}$, that is,
$$
E(X)=tP_{\omega_1}(X)+(1-t)P_{\omega_2}(X),\quad X\in\bor{\Omega},
$$
for some $t\in(\frac{1}{2},1)$. For any $X\in\bor{\Omega}$,
the projections $P_{\omega_1}(X)$ and $P_{\omega_2}(X)$ commute, so
that also their product $Q:=P_{\omega_1}(X)P_{\omega_2}(X)$ is
projection. The differences $P_{\omega_1}(X)-Q$ and
$P_{\omega_1}(X)-Q$ are mutually orthogonal projections, and the
spectral decomposition of the effect $E(X)$ is
$$
E(X)=t(P_{\omega_1}(X)-Q)+(1-t)(P_{\omega_2}(X)-Q)+Q.
$$
Assume that the spectrum of $E(X)$ lies under $\frac{1}{2}$. This
means that $P_{\omega_1}(X)-Q=Q=O$, implying also that
$P_{\omega_1}(X)=O$ and by equation (\ref{nolla}),
$\mu_{\Omega}(X\omega_1^{-1})=0$. In the right translations the
sets of $\mu_{\Omega}$-measure zero remain $\mu_{\Omega}$-measure
zero. Hence, $\mu_{\Omega}(X\omega_2^{-1})=0$, and by equation
(\ref{nolla}), $P_{\omega_2}(X)=O$. Therefore, the spectrum of any
$E(X)\neq O$ extends above $\frac{1}{2}$. If $O\neq E(X)\neq I$,
then also the complement effect $E(X')=I-E(X)$ has $t$ or $1$ in
spectrum. This implies that either $0$ or $1-t$ belongs in the
spectrum of $E(X)$, showing that the observable $E$ is regular.
\end{example}

Even though there are regular fuzzy observables, regularity is a
surprisingly strong condition. Proposition \ref{reg:con} shows
that many typical fuzzy observables (see Section \ref{locR}) are
not regular.

\begin{proposition}\label{reg:con}
Let $\Omega$ be non-discrete and $\rho:\bor{\Omega}\to [0,1]$ a
probability measure. If $\rho$ is absolutely continuous with
respect to the Haar measure $\mu_{\Omega}$, then the fuzzy
observable $E_{\rho}$ is not regular.
\end{proposition}

\begin{proof}
As in the proof of Proposition \ref{norm}, we get the inequality
\begin{equation}\label{rhosup}
\no{E_{\rho}(X)}\leq \sup_{\omega\in\Omega}\rho(\omega X)
\end{equation}
for any $X\in\bor{\Omega}$. 
For every $\epsilon >0$ there exists a $\delta_{\epsilon} >0$ such that
$\rho(\omega X)<\epsilon$ whenever $\mu_{\Omega}(\omega
X)<\delta_{\epsilon}$ (see \cite[Theorem 3.5]{Fol1}). Hence,
$\mu_{\Omega}(X)<\delta_{\epsilon}$ implies that $\rho(\omega X)<\epsilon$
for every $\omega\in\Omega$.  
Since $\Omega$ is not discrete, the Haar measure $\mu_{\Omega}$
has arbitrarily small positive values. Thus we can choose
$X_1\in\bor{\Omega}$ such that 
$0<\mu_{\Omega}(X_1)<\delta_{\frac{1}{2}}$. Now  $\rho(\omega
  X)<\frac{1}{2}$ for every $\omega\in\Omega$ and this together with
  equation (\ref{rhosup}) gives
$\no{E_{\rho}(X_1)}<\frac{1}{2}$.  Moreover, $E_{\rho}(X_1)\neq O$ by equation (\ref{nolla}).

\end{proof}

\subsection{Informational equivalence}\label{equivalence}

We shall next investigate the state distinction power of fuzzy
observables. In general, two observables $E_1,E_2$ are called
\emph{informationally equivalent} if for any two states $S,T\in\trh_1^+$,
$$
V_{E_1}(S)=V_{E_1}(T)\ \Leftrightarrow\
V_{E_2}(S)=V_{E_2}(T).
$$
This means that informationally equivalent observables have the same
ability to distinguish between different states. Clearly, $E_1$ and $E_2$ are informationally equivalent if and only if for any two unit vectors
$\varphi,\psi\in\hi$ we have
$$
p_{\varphi}^{E_1}=p_{\psi}^{E_1}\ \Leftrightarrow\
p_{\varphi}^{E_2}=p_{\psi}^{E_2}.
$$

Since a fuzzy observable is a coarse-graining of a sharp observable,
its ability to distinguish states does not exceed that of the sharp
observable. However, a fuzzy observable can be informationally equivalent to
the underlying sharp one. Proposition \ref{infor} deals with this matter.

Let the group $\Omega=G/H$ be Abelian. If $\rho$ is a probability
measure on $\Omega$, its Fourier-Stieltjes transform $\hat{\rho}$
is the bounded continuous function on the dual group
$\hat{\Omega}$, defined by
$$
\hat{\rho}(\xi)=\int_{\Omega}\overline{\dual{\omega}{\xi}}\ud\rho(\omega)=
\int_{\Omega}\dual{\omega}{\xi^{-1}}\ud\rho(\omega),
$$
where $\dual{\omega}{\xi}=\xi(\omega)$ is the value of the
character $\xi$ at the point $\omega$.

\begin{proposition}\label{infor}
The observables $P$ and $E_{\rho}$ are informationally equivalent
if and only if $\supp(\hat{\rho})=\hat{\Omega}$.
\end{proposition}

\begin{proof}
Assume that the latter condition is fulfilled. Let
$\varphi,\psi\in\hi$ be two unit vectors such that
$p_{\varphi}^{E_\rho}=p_{\psi}^{E_\rho}$. Using equation (\ref{conv1})
and taking Fourier-Stieltjes transforms we get
$\widehat{p_{\varphi}^P}(\xi)\widehat{\rho}(\xi)=\widehat{p_{\psi}^P}(\xi)\widehat{\rho}(\xi)$
for all $\xi\in\hat{\Omega}$. This implies that
$\widehat{p_{\varphi}^P}(\xi)=\widehat{p_{\psi}^P}(\xi)$ in a dense
set. Since the functions $\widehat{p_{\varphi}}$ and
$\widehat{p_{\psi}^P}$ are continuous, it follows that
$\widehat{p_{\varphi}^P}=\widehat{p_{\psi}^P}$, and so
$p_{\varphi}^P=p_{\psi}^P$. Thus the observables $P$ and $E_{\rho}$
are informationally equivalent.

Assume now that $\supp(\hat{\rho})\varsubsetneqq\hat{\Omega}$. We
have to show that there are unit vectors $\psi,\varphi\in L^2(\Omega,\mu_{\Omega},\K)$ such that
$p_{\psi}^P\neq p_{\varphi}^P$ but $p^{E_\rho}_{\psi}=p^{E_\rho}_{\psi}$.
Let us denote by $U$ the complement of $\supp(\hat{\rho})$. Fix
$\hat{\omega}\in\hat{\Omega}$ such that $\hat{\omega}^{-1}$
belongs to $U$. Then the set $\hat{\omega}U$ is an open
neighborhood of the neutral element $e$ and there is a symmetric
open set $V$ such that $e\in V$ and $VV\subset \hat{\omega}U$.
According to Urysohn's lemma there exists a continuous function
$f:\hat{\Omega}\to [0,1], f\neq 0,$ having compact support such
that $\supp(f)\subset V$. The convolution $f\ast f^{\ast}$ is a
positive definite continuous function with compact support
contained in $\hat{\omega}U$. By the Fourier inversion theorem
there is a positive function $F'\in L^1(\Omega)$ such that
$\hat{F'}=f\ast f^*$. The product $F:=\hat{\omega}^{-1}F'$ is in
$L^1(\Omega)$ and
$$
\hat{F}(\xi)=(f\ast f^*)(\hat{\omega}\xi).
$$
It follows that the support of $\hat{F}$ is contained in $U$ and
therefore $\hat{\rho}\hat{F}=0$. This means that
\begin{equation}\label{info:t1}
\rho\ast F=0.
\end{equation}
We now write $F=F_1 - F_2 + iF_3 - iF_4$, where the
$F_j,j=1,2,3,4,$ are nonnegative functions and belong to
$L^1(\Omega)$. Since $F\neq 0$, either $F_1\neq F_2$ or $F_3\neq
F_4$. We consider the case $F_1\neq F_2$, the other one being
similar.

Equation (\ref{info:t1}) implies that
\begin{equation}\label{info:t2}
\rho\ast F_1=\rho\ast F_2.
\end{equation}
As $\no{\rho\ast F_1}_1=\no{F_1}_1$ and similarly $\no{\rho\ast
F_2}_1=\no{F_2}_1$, we have $\no{F_1}_1=\no{F_2}_1$. Therefore we
may normalize $\no{F_1}_1=\no{F_2}_1=1$ keeping equation
(\ref{info:t2}) valid. Fix a unit vector $\vartheta\in\K$ and
denote $\psi(\omega)=\sqrt{F_1(\omega)}\vartheta$ and
$\varphi(\omega)=\sqrt{F_2(\omega)}\vartheta$ for all
$\omega\in\Omega$. Then $\psi,\varphi\in
L^2(\Omega,\mu_{\Omega},\K)$ are states, $p_{\psi}^P\neq
p_{\varphi}^P$ and $p^{E_\rho}_{\psi}=p^{E_\rho}_{\psi}$.
\end{proof}

\begin{remark}
In \cite{AliDoe} the Euclidean covariant localization observables
were studied. If $E$ is a Euclidean covariant localization
observable, then it is in particular covariant under translations.
For a translationally covariant fuzzy observable Proposition
\ref{infor} is applicable. Our result therefore shows that in
\cite[Theorem 2]{AliDoe} one needs the additional assumption that the
Fourier-Stieltjes transform of the probability measure in question has
 all of $\R^3$ as its support. The same observation applies e.g. to
\cite[p. 124]{OQP} where \cite{AliDoe} is quoted. 
\end{remark}

\section{Examples}\label{examples}

In this section we illustrate the concepts and results of the
previous sections by the Stern-Gerlach experiment and by some
localization observables.

\subsection{The Stern-Gerlach experiment}
In the most simplified, though prototypical textbook description
of a Stern-Gerlach experiment a spin-$\frac 12$ atom  prepared in
a vector state $\phi_{{\rm orbit}}\otimes\psi_{{\rm spin}}$ is
directed (say, horizontally) to a (say,
vertical) Stern-Gerlach magnetic field. Writing $\psi_{{\rm
spin}}=c_{\uparrow}\psi_{\uparrow}+c_{\downarrow}\psi_{\downarrow}$,
the action of the magnet on the atom is described by a unitary
transformation coupling its spin and spatial degrees of freedom:
$$
\phi_{{\rm orbit}}\otimes \psi_{{\rm spin}} \mapsto c_{\uparrow}\phi_{\uparrow}\otimes
\psi_{\uparrow}+c_{\downarrow}\phi_{\downarrow}\otimes
\psi_{\downarrow}.
$$
The deflected atom hits a screen producing a spot on it. The
counting of the spots (that is, the collecting of the measurement
statistics of up and down deflected atoms) can be modelled by
projection operators $P_{\uparrow}$ and $P_{\downarrow}$
representing  the localization of the atom in the upper or lower
half planes of the screen. The relevant measurement outcome
probabilities are then
\begin{eqnarray*}
p_{\uparrow} &=&
|c_{\uparrow}|^2\ip{\phi_{\uparrow}}{P_{\uparrow}\phi_{\uparrow}}+
|c_{\downarrow}|^2\ip{\phi_{\downarrow}}{P_{\uparrow}\phi_{\downarrow}},\\
p_{\downarrow} &=&
|c_{\uparrow}|^2\ip{\phi_{\uparrow}}{P_{\downarrow}\phi_{\uparrow}}+
|c_{\downarrow}|^2\ip{\phi_{\downarrow}}{P_{\downarrow}\phi_{\downarrow}}.
\end{eqnarray*}
When these probabilities are expressed in terms of a two-valued
spin observable $E$ and the incoming spin state $\psi_{{\rm
spin}}$, that is,
 $\ip{\psi_{{\rm spin}}}{E_{\uparrow}\psi_{{\rm spin}}}:=p_{\uparrow}$ and
$\ip{\psi_{{\rm spin}}}{E_{\downarrow}\psi_{{\rm
spin}}}:=p_{\downarrow}$, the measured spin observable $E$ has the
form
\begin{eqnarray*}
E_{\uparrow} &=&
\ip{\phi_{\uparrow}}{P_{\uparrow}\phi_{\uparrow}}P[\psi_{\uparrow}]+
\ip{\phi_{\downarrow}}{P_{\uparrow}\phi_{\downarrow}}P[\psi_{\downarrow}],\\
E_{\downarrow} &=&
\ip{\phi_{\uparrow}}{P_{\downarrow}\phi_{\uparrow}}P[\psi_{\uparrow}]+
\ip{\phi_{\downarrow}}{P_{\downarrow}\phi_{\downarrow}}P[\psi_{\downarrow}].
\end{eqnarray*}
If $\ip{\phi_{\uparrow}}{P_{\uparrow}\phi_{\uparrow}}=
\ip{\phi_{\downarrow}}{P_{\downarrow}\phi_{\downarrow}}=1$ and
$\ip{\phi_{\downarrow}}{P_{\uparrow}\phi_{\downarrow}}=
\ip{\phi_{\uparrow}}{P_{\downarrow}\phi_{\uparrow}}=0$, we get the
sharp spin observable $s_{\uparrow}$ with the spectral projections
$E_{\uparrow}=P[\psi_{\uparrow}]$ and
$E_{\downarrow}=P[\psi_{\downarrow}]$. In general, the measured
spin observable $E$ is a fuzzy observable with respect to
$s_{\uparrow}$. Apart from the trivial observable
$E_{\uparrow}=E_{\downarrow}=\frac{1}{2}I$, the measured fuzzy
spin observables are informationally equivalent to the sharp one,
$s_{\uparrow}$. Being a two-valued observable $E$ is regular. The
norms of $E_{\uparrow}$ and $E_{\downarrow}$ are their larger
eigenvalues which are equal to one only when these effects are the
spin projections. (A more realistic account of a Stern-Gerlach
experiment as a spin measurement experiment can be read, for
instance, in \cite{OQP}, but the basic picture is essentially the same.)

\subsection{Localization on $\R$}\label{locR}

The usual position observable on $\R$ is the spectral measure
$P:\bor{\R}\to \mathcal{L}(L^2(\R))$, $P(X)\psi=\chi_X\psi$. It is
covariant under the regular representation $U:\R\to\uh$,
$[U(x)\psi](y)=\psi(y-x)$. The representation $U$ is induced from
the trivial representation of the trivial subgroup $\{0\}$. All
$\R$-covariant localization observables are characterized explicitly
in \cite{CDT}. We emphasize that there exist also other covariant
position observables than the sharp and fuzzy localization
observables. Even non-commutative covariant localization observables
exist.

Davies \cite{Davies76} calls an observable
$E_f:\bor{\R}\to\mathcal{L}(L^2(\R))$, $E_f(X)=(f\ast\chi_X)(P)$,
where $f$ is a probability density function, an approximate
position observable. This is a covariant fuzzy observable, the
probability measure being $\check{f}\ud\mu$, where
$\check{f}(x)=f(-x)$ (see Example \ref{covar:ex}). According to Proposition \ref{infor} the
approximate position observable $E_f$ is informationally
equivalent to the sharp position observable $P$ if and only if
$\supp(\hat{f})=\R$. For example, if $f$ is a Gaussian function,
then the Fourier transform of $f$ is also Gaussian and $E_f$ is
informationally equivalent to $P$. Also the functions
$\frac{1}{2a}\chi_{[-a,a]},\ a>0,$ yield approximate position
observables that are informationally equivalent to $P$. An example
of an approximate position observable which is not informationally
equivalent to $P$, is obtained if one takes $f(x)=\sin^2(2\pi
x)/2\pi^2x^2$. In this case $\hat{f}(\xi)=\max
(1-\frac{1}{2}|\xi|,0)$ and $\supp(\hat{f})=[-2,2]$.

\subsection{Localization on $\torus$}

Let us denote by $\torus$ the set of all complex numbers modulus one. It is a compact Abelian group and the Haar measure
$\mu=\mu_{\torus}$ of $\torus$ is just the normalized and
translated Lebesgue measure. By a $\torus$-covariant localization
observable we mean an observable $E:\bor{\torus}\to
\mathcal{L}(L^2(\torus))$ satisfying the covariance condition
$$
U(a)E(X)U(a)^*=E(aX),
$$
where $[U(a)\psi](z)=\psi(a^{-1}z)$.

The family $\{e_k\}_{k\in\Z}$ of functions
$$
e_k:\torus\to\C,\ e_k(a)=a^k,
$$
is an orthonormal basis for $L^2(\torus)$. Theorem 1 of
\cite{CDLP} shows that any $\torus$-covariant localization
observable $E$ has the form
\begin{equation}\label{tloc}
E(X)=\sum_{n,m\in\Z} c_{n,m} \int_X z^{m-n}\ud\mu(z)\
\ketbra{e_n}{e_m},\quad X\in\boto,
\end{equation}
where the numbers $c_{n,m}$ satisfy the following two conditions:
\begin{itemize}
\item[(a)] $c_{n,n}=1$ for all $n\in\Z$;
\item[(b)] $\sum_{n,m=-k}^{k} c_{n,m}\ketbra{e_n}{e_m}\geq O$ for
all $k\in\N$.
\end{itemize}
Note that the canonical spectral measure $P$ corresponds to the
matrix $(c_{n,m})$, where $c_{n,m}=1$ for all $n,m\in\Z$.

This simple structure of $\torus$-covariant localization
observables gives us a possibility to investigate further the
fuzzy observables on $\torus$. We say that the observables $E_1$
and $E_2$ \emph{commute} if $E_1(X)E_2(Y)=E_2(Y)E_1(X)$ for all
$X,Y\in\boto$. An observable $E$ is \emph{commutative} if it
commutes with itself.

\begin{lemma}\label{com1}
A $\torus$-covariant localization observable $E$ corresponding to
the matrix $(c_{n,m})$ commutes with $P$ if and only if
\begin{equation}\label{eq:com2}
c_{n+k,m+k}=c_{n,m}
\end{equation}
for all $n,m,k \in\Z$.
\end{lemma}

\begin{proof}
The proof is similar to the proof of Proposition 1 in \cite{CDLP},
which characterizes commutative $\torus$-covariant localization
observables. Define
$$
\mu_{n,m,Y}:= \langle e_n | P(X)E(Y)-E(Y)P(X)| e_m \rangle
$$
for all $n,m\in \Z$ and $X,Y \in \boto$. Now
$$
\langle e_n | P(X)E(Y) | e_m \rangle = \sum_{s=-\infty}^{\infty}
c_{s,m} \int_X z^{s-n}\ \ud \mu(z) \int_Y z^{m-s}\ \ud \mu(z)
$$
and
$$
\langle e_n | E(Y)P(X) | e_m \rangle = \sum_{s=-\infty}^{\infty}
c_{n,s} \int_X z^{m-s}\ \ud \mu(z) \int_Y z^{s-n}\ \ud \mu(z).
$$
It follows that for all $k\in \Z$,
$$
\int_{\torus} z^{-k}\ \ud \mu_{n,m,Y}(z) = [c_{n+k,m} - c_{n,m-k}]
\int_Y z^{m-n-k}\ \ud \mu(z).
$$
If $P(X)E(Y)=E(Y)P(X)$ for all $X,Y \in \boto$, equation
(\ref{eq:com2}) is satisfied. Also, if (\ref{eq:com2}) is valid,
then
\begin{eqnarray*}
\mu_{n,m,Y}(X) &=& \sum_{k=-\infty}^{\infty} [c_{n+k,m}-c_{n,m-k}] \int_X z^k\ \ud \mu(z) \int_Y z^{m-n-k}\ \ud \mu(z) \\
&=& 0
\end{eqnarray*}
for all $n,m\in \Z$.
\end{proof}

\begin{proposition}\label{com2}
A $\torus$-covariant localization observable $E$ is fuzzy with
respect to $P$ if and only if $E$ commutes with $P$.
\end{proposition}

\begin{proof}
It is clear that if $E$ is fuzzy with respect to $P$, then it also
commutes with $P$. Let now $E$ be a $\torus$-covariant
localization observable that commutes with $P$. Then according to
Lemma \ref{com1}, the matrix elements $c_{n,m}$ of $E$ satisfy
$c_{n+k,m+k}=c_{n,m}$ for all $n,m,k\in\Z$. Let us define the function
$\Phi:\Z\to\C$ by $\Phi(k)=c_{k,0}$. Since $c_{n,m}=c_{n-m,0}$
for all $n,m\in\Z$, the function $\Phi$ determines all the
matrix elements $c_{n,m}$. Moreover, $\Phi$ has following
properties:
\begin{itemize}
\item[(i)] $\Phi(0)=1$;
\item[(ii)] $\Phi(-k)=\overline{\Phi(k)}$ for all $k\in\Z$;
\item[(iii)] $|\Phi(k)|\leq 1$ for all $k\in\Z$;
\item[(iv)] $\sum_{n,m=-N}^N \bar{d}_n \Phi(n-m) d_m \geq 0$ for all sequences $(d_n)_{n\in\Z}\subset\C$
and every $N\in\N$.
\end{itemize}
Condition (iv) means that $\Phi$ is positive definite. Herglotz's
theorem \cite{Herglotz} (see e.g. \cite[p. 95]{Fol2}, \cite[p.
293]{HR}) states that in this situation there is a unique positive
measure $\rho\in\MT$ such that $\Phi$ is given by
\begin{equation}
\Phi(k)=\int_{\torus} \langle k,z \rangle \ud \rho(z) =
\int_{\torus} z^k\ud \rho(z).
\end{equation}
Because
$$
\rho(\torus)=\int_{\torus}\ud\rho(z)=\Phi(0)=1,
$$
$\rho$ is a probability measure. Now
\begin{eqnarray*}
\ip{e_n}{E(X)|e_m} &=& \Phi(n-m)\int_X z^{m-n}\ud\mu(z) \\
&=& \int_{\torus}y^{n-m}\ud\rho(y)\  \int_{\torus}\chi_X(z)z^{m-n}\ud\mu(z) \\
&=& \int_{\torus}\int_{\torus}(y^{-1}z)^{m-n}\chi_X(z)\ud\mu(z)\ud\rho(y)\\
&=& \int_{\torus}\int_{\torus}z^{m-n}\chi_X(yz)\ud\mu(z)\ud\rho(y) \\
&=& \int_{\torus}z^{m-n}\left( \int_{\torus}
\chi_{z^{-1}X}(y)\ud\rho(y) \right) \ud\mu(z)\\
&=& \int_{\torus}z^{m-n}\rho(z^{-1}X)\ud\mu(z),
\end{eqnarray*}
so $E$ can be written as
\begin{equation}\label{can1}
E(X)=\int_{\torus}\rho(z^{-1}X)\ud P(z).
\end{equation}
\end{proof}

We finally note that not all commutative $\torus$-covariant
localization observables are fuzzy. This can be demonstrated by
the following example, due to A. Toigo \cite{Toigo}. Set
$c_{n,n}=1$ for all $n\in\Z$ and for all $n,m\in\Z, n\neq m$
define
$$
c_{n,m}=\left\{ \begin{array}{ll} 1, & \textrm{if n and m are even;} \\
0, & \textrm{otherwise.}
\end{array} \right.
$$

Then $(c_{n,m})$ satisfies conditions (a) and (b) stated after eq. (\ref{tloc}), and using
Proposition 1 in \cite{CDLP} it is easy to see that the
localization observable $E$ defined by $(c_{n,m})$ is commutative.
Since, for example, $c_{2,4}=1$ and $c_{3,5}=0$, Lemma \ref{com1}
shows that $E$ does not commute with $P$ and therefore, is not
fuzzy.

\section*{Acknowledgments}

We thank Alessandro Toigo for his useful comments on an early
version of this paper.


\end{document}